%% file: main.tex
\documentclass[11pt,a4paper]{article}

\usepackage[utf8]{inputenc}
\usepackage[T1]{fontenc}
\usepackage{amsmath,amssymb,amsthm}
\usepackage{graphicx}
\usepackage{hyperref}
\usepackage{booktabs}
\usepackage{tikz}
\usetikzlibrary{shapes,arrows,positioning,fit,backgrounds}
\usepackage{algorithm}
\usepackage{algpseudocode}
\usepackage{xcolor}
\usepackage{geometry}
\newtheorem{definition}{Definition}

\title{\textbf{Distributed Security: From Isolated Properties to Synergistic Trust}}
\author{
    Minghui Xu$^{1,2}$
}
\date{
    $^1$Shandong University, China \\
    $^2$Quan Cheng Laboratory, China \\
    \texttt{mhxu@sdu.edu.cn}
}

\begin{document}

\maketitle

\begin{abstract}
Over the past four decades, distributed security has undergone a remarkable transformation---from crash-fault tolerant protocols designed for controlled environments to sophisticated Byzantine-resilient architectures operating in open, adversarial settings. This vision paper examines this evolution and argues for a fundamental shift in how we approach distributed security: from studying individual security properties in isolation to understanding their synergistic combinations. We begin by conclude four foundational properties, \textit{agreement, consistency, privacy, verifiability, accountability}. We trace their theoretical origins and practical maturation. We then demonstrate how the frontier of research now lies at the intersection of these properties, where their fusion creates capabilities that neither property could achieve alone. Looking forward, we identify critical research challenges: discovering new security properties driven by emerging applications, developing systematic frameworks for property convergence, managing the computational overhead of cryptographic primitives in high-performance consensus layers, and addressing post-quantum and human-factor challenges. The future of distributed security lies not in improving individual properties, but in understanding and harnessing their synergies to build a singular fabric of trust.
\end{abstract}

\tableofcontents
\newpage

\input{sections/introduction}
\input{sections/distributed-security}
\input{sections/key-properties}
\input{sections/frontier-fusions}
\input{sections/future-challenges}
\input{sections/conclusion}

\bibliographystyle{plain}
\bibliography{references}

\end{document}

%% file: sections/introduction.tex
\section{Introduction}

\subsection{The Transition of Trust}

The evolution of distributed systems over the past five decades represents one of the most significant transformations in computing history. What began as simple mechanisms for crash-fault tolerance in closed, trusted networks has evolved into sophisticated Byzantine-resilient architectures capable of operating in open, adversarial environments. This transition reflects not merely a technical progression, but a fundamental reconceptualization of how \textbf{trust} is established, maintained, and verified in systems that span organizational boundaries, jurisdictional domains, and competing interests.

In the early days of distributed computing, systems were designed to operate within controlled environments where failures could be modeled as probabilistic events rather than strategic attacks. The primary concern was ensuring continued operation despite node crashes, network partitions, or storage failures. Under these assumptions, protocols could be designed with relatively simple failure models: nodes were either working correctly or completely failed, and the network either delivered messages or did not. This era gave rise to foundational protocols such as two-phase commit, viewstamped replication, and eventually Paxos~\cite{lamport1998paxos}, which provided strong guarantees about system behavior under crash failures.

However, as distributed systems expanded beyond controlled data centers into the broader internet, financial systems, and critical infrastructure, the threat model fundamentally changed. The adversaries facing modern distributed systems are no longer abstract probabilities but intelligent, strategic actors capable of arbitrary behavior. These adversaries may control compromised nodes within the system, manipulate network communications, or even operate their own nodes with malicious intent from the outset. The rise of cryptocurrencies, decentralized finance (DeFi), and cross-organizational computation has made this adversarial model not merely theoretical but a practical reality that system designers must address.

This shift from crash-fault tolerance to Byzantine fault tolerance marks a fundamental transformation in how we conceptualize trust in distributed systems. No longer can we rely on the assumption that the majority of participants will behave correctly simply because they have no incentive to misbehave. Instead, we must design systems that remain secure even when faced with coordinated attacks from sophisticated adversaries who may have significant financial or strategic incentives to compromise system integrity. This new reality demands a fundamental rethinking of distributed security architecture.

\subsection{The Motivation}

Traditional security boundaries were designed for a world where systems could be clearly delineated into trusted and untrusted zones. Firewalls established perimeters around organizational networks. Access control lists defined who could access which resources. Trusted computing bases provided secure enclaves for sensitive operations. These mechanisms all shared a common assumption: that the primary security challenge was preventing unauthorized access to a well-defined trusted domain.

The modern distributed landscape has shattered these assumptions in multiple dimensions. Consider the challenges facing contemporary distributed applications. First, modern systems often span multiple administrative domains, each with their own security policies, trust assumptions, and regulatory requirements. A supply chain management system may involve manufacturers, logistics providers, retailers, and financial institutions, none of whom fully trust each other yet all of whom must cooperate to maintain accurate records. Second, participants in decentralized systems may have directly conflicting interests. In a financial exchange, buyers benefit from lower prices while sellers benefit from higher ones; in a prediction market, participants may have incentives to manipulate outcomes. Third, privacy regulations such as GDPR, HIPAA, and various national data protection laws demand that computation occur without exposing sensitive personal information, even when that computation involves multiple parties. Fourth, users increasingly demand cryptographic proof that systems behaved correctly, rather than simply trusting service providers to act appropriately.

\textbf{These challenges cannot be addressed by any single security property in isolation}. A system might achieve perfect consensus on transaction ordering yet leak private information about transaction contents to observing adversaries. It might preserve complete privacy of individual inputs yet provide no mechanism for participants to verify that the computation was performed correctly. It might generate impeccable cryptographic proofs of correctness yet be unable to reach agreement when some participants behave maliciously. The insufficiency of isolated properties becomes particularly apparent when we consider real-world applications.

Consider a decentralized financial system that enables users to trade assets without relying on a central exchange. Such a system must \textbf{simultaneously} achieve agreement among distributed nodes on the order of trades, ensure that account balances remain consistent and cannot be spent twice, protect the privacy of individual trading strategies and positions, and provide cryptographic proof that all operations were executed according to the agreed-upon rules. If any one of these properties fails, the system becomes either unusable, insecure, or both. This is why a \emph{multi-dimensional approach} is essential.

We identify five key properties that together form the foundation of distributed security. \emph{Agreement} enables distributed nodes to reach consensus on a single version of truth, even in the presence of malicious actors who may attempt to create conflicting views of system state. \emph{Consistency} guarantees that once agreement is reached, data remains uniform and persistent across all participants, preventing divergence that could enable double-spending or other inconsistencies. \emph{Privacy} enables computation over distributed inputs without revealing the underlying sensitive data to any participant or observer. \emph{Verifiability} provides the ability to cryptographically prove that computations were executed correctly, enabling trust without requiring trust in any individual component. \emph{Accountability} ensures that misbehavior can be detected and attributed to specific parties through cryptographic evidence, thereby enabling deterrence, auditability, and enforceable consequences in distributed systems.

\subsection{Toward Synergistic Architecture}

Historically, each of these properties has been studied and implemented in relative isolation by different research communities using different formalisms and techniques. Consensus researchers developed increasingly sophisticated protocols for agreement, from early two-phase commit protocols through Paxos~\cite{lamport1998paxos} and Raft~\cite{ongaro2014raft} to modern Byzantine fault tolerant protocols, but typically assumed that privacy and verifiability were concerns to be addressed by other layers of the system. Database researchers focused on consistency models and transaction processing, developing powerful techniques for maintaining data integrity, but often assumed crash-fault models that fail to capture adversarial behavior. Cryptographers developed elegant protocols for privacy and verifiability, including secure multi-party computation~\cite{yao1982protocols} and zero-knowledge proofs~\cite{goldwasser1989knowledge}, but often assumed communication and agreement primitives that are impractical in real distributed deployments.

This siloed approach was productive in its time, enabling significant advances in each individual area. However, it is \textbf{increasingly untenable} for modern applications that demand simultaneous guarantees across multiple properties. The frontier of distributed security research now lies at the \textbf{intersection} of these properties, where their combination creates new capabilities that neither property could provide alone. We are witnessing the emergence of protocols that fuse agreement with consistency to create blockchains~\cite{nakamoto2008bitcoin}, verifiability with consistency to enable zero-knowledge rollups, privacy with verifiability to enable collaborative cryptographic protocols~\cite{ozdemir2022collaborative}, and many other combinations that blur the traditional boundaries between research areas.

This paper argues that distributed security is shifting from a collection of isolated protocols into a \textbf{unified, synergistic architecture} where properties are no longer independent silos but interacting components of a coherent system design. The future of distributed security lies not in improving individual properties in isolation, but in understanding how these properties can be combined, traded off against each other when necessary, and made to reinforce each other when possible. 

%% file: sections/distributed-security.tex
\section{Distributed Security}

\subsection{The Brief History of Distributed Security}

The field of distributed security has evolved through several distinct eras, each characterized by different threat models, technical capabilities, and application domains. Understanding this evolution provides essential context for appreciating the current state of the field and the challenges that motivate the fusions we discuss in later sections.

The \textbf{foundation era}, spanning roughly from the 1970s through the 1990s, established the theoretical vocabulary and formal frameworks that continue to underpin the field today. Lamport's seminal 1977 work~\cite{lamport1977proving} on safety and liveness properties formalized the distinction between guarantees that ``nothing bad ever happens'' (safety) and guarantees that ``something good eventually happens'' (liveness). This seemingly simple distinction has profound implications: safety properties can be violated in finite time and must be maintained at every step, while liveness properties can only be violated over infinite time and depend critically on fairness assumptions about the system environment. The Byzantine Generals Problem~\cite{lamport1982byzantine}, published in 1982, formalized the challenge of reaching consensus when some participants may behave maliciously, establishing fundamental limits that continue to shape protocol design. The paper demonstrated that in a synchronous system with message authentication, at least $3f+1$ nodes are necessary and sufficient to tolerate $f$ Byzantine failures---a result that might seem surprising until one realizes that with only $3f$ nodes, the Byzantine nodes could potentially present different values to different honest nodes, making it impossible for honest nodes to distinguish truth from fabrication.

The theoretical exploration continued with impossibility results that, paradoxically, guided practical protocol design by clarifying what could and could not be achieved. The FLP impossibility result~\cite{fischer1985impossibility}, published in 1985, demonstrated that deterministic consensus is impossible in asynchronous systems with even a single crash failure. The proof hinges on the possibility that the system can remain forever undecided, with messages between crucial nodes delayed just long enough to prevent consensus. Rather than ending the quest for practical consensus, this result focused research on \emph{partially synchronous models} and randomized protocols that could achieve consensus with probability approaching one. These theoretical foundations, while abstract, established the conceptual framework within which all subsequent distributed security research operates.

The \textbf{practical era}, spanning roughly the 1990s through the 2010s, marked the transition from theoretical constructs to real-world deployment. Practical Byzantine Fault Tolerance (PBFT)~\cite{castro1999pbft}, published in 1999, was a watershed moment that demonstrated Byzantine-resilient systems could achieve performance comparable to crash-fault tolerant systems under realistic conditions. PBFT introduced a three-phase protocol---pre-prepare, prepare, and commit---that ensures all honest nodes agree on the same sequence of operations even when up to one-third of nodes behave arbitrarily. While its $O(n^2)$ message complexity limited scalability for large networks, PBFT proved that Byzantine fault tolerance was not merely a theoretical curiosity but a practical engineering concern. The protocol has since been implemented in numerous production systems and remains influential in the design of modern consensus protocols.

This era also witnessed the maturation of distributed database technology. Systems like Google's Spanner~\cite{corbett2012spanner} demonstrated that globally-consistent distributed transactions were achievable at scale through innovations such as atomic clock synchronization, which provides a global notion of time that can be used to order transactions across geographically distributed data centers. The Raft consensus algorithm~\cite{ongaro2014raft}, designed explicitly for understandability, made distributed consensus techniques accessible to a broader engineering audience and has been widely adopted in production systems. These developments showed that the theoretical insights from the foundation era could be translated into practical, reliable systems that meet the demands of real-world applications.

The \textbf{decentralization era}, beginning in 2009 with the launch of Bitcoin~\cite{nakamoto2008bitcoin}, fundamentally changed the landscape of distributed security. For the first time, a Byzantine-fault tolerant system operated at global scale without any trusted authority. Bitcoin's proof-of-work consensus mechanism introduced a radically different approach to achieving agreement: rather than relying on known identities and message passing among a fixed set of nodes, Bitcoin's security derives from computational investment. Miners compete to solve cryptographic puzzles, and the longest chain of valid blocks represents the consensus view of transaction history. This mechanism, while energy-intensive, demonstrated that \emph{decentralized agreement} was possible in permissionless settings where participants could join and leave freely without registration or authorization.

Ethereum~\cite{wood2014ethereum}, launched in 2015, extended Bitcoin's vision by introducing programmable smart contracts---general-purpose code that executes on the blockchain, enabling complex decentralized applications beyond simple value transfer. The subsequent development of proof-of-stake mechanisms~\cite{buterin2020ethereum}, which Ethereum ultimately adopted, replaced computational work with economic stake as the basis for security, dramatically improving energy efficiency while maintaining strong security guarantees under appropriate economic assumptions. This era represents the culmination of decades of research: systems that provide meaningful security guarantees without relying on any trusted entity, operated by anonymous participants who may have no prior relationship and no reason to trust each other, yet can nevertheless cooperate to maintain a consistent view of shared state.

\begin{figure}[htbp]
\centering
\includegraphics[width=\textwidth]{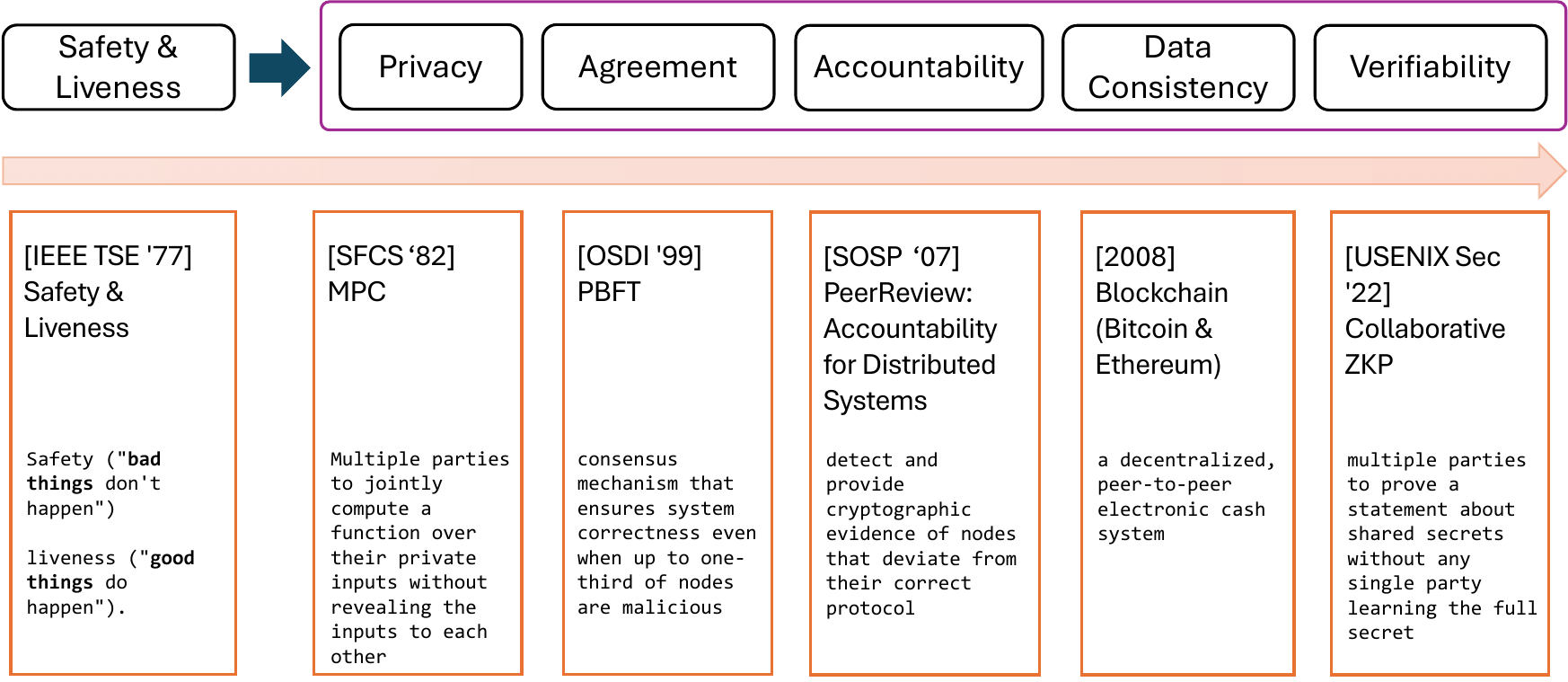}
\caption{The Evolution of Distributed Security: A Forty-Five Year Journey}
\label{fig:timeline}
\end{figure}

Figure~\ref{fig:timeline} illustrates this evolution, showing how the field has progressed from theoretical foundations through practical deployment to the current era of decentralization. Each era built upon the insights of its predecessors, with theoretical results guiding practical design and practical experience revealing new theoretical questions. The timeline highlights the key milestones that have shaped our current understanding of distributed security.

\subsection{How to Analyze Distributed Security}

Understanding and designing secure distributed systems requires a \textbf{systematic analytical framework} that moves beyond ad hoc reasoning to principled analysis. We propose a methodology comprising six interrelated components: problem definition, threat modeling, security goals, system assumptions, technique selection, and evaluation methods. This framework enables rigorous analysis of existing systems and principled design of new ones.

\emph{Problem definition} is the foundation upon which all subsequent analysis rests. Many system failures can be traced not to implementation bugs but to poorly defined problems where the fundamental service being provided, the participants and their roles, and the correctness conditions remain unclear or ambiguous. A well-defined problem specification must answer several questions precisely: What is the fundamental service being provided? Who are the participants, and what are their roles? What are the safety conditions that must never be violated? What are the liveness conditions that must eventually be satisfied? How do these conditions interact with system performance, scalability, and usability requirements? A precise problem definition enables meaningful comparison of different solutions and provides the basis for rigorous correctness proofs.

The \emph{threat model} defines what adversaries can do within the system model. This includes the corruption model, which specifies whether adversaries can corrupt participants, how many they can corrupt, and what type of corruption is possible (crash failures, Byzantine failures, or something in between). Network assumptions specify what control the adversary has over communication: can they observe messages, delay them, reorder them, drop them, or inject fabricated messages? Computational assumptions specify whether the adversary is computationally bounded, and if so, what cryptographic hardness assumptions are considered valid. Adaptivity considerations specify whether the adversary can adaptively corrupt participants based on observed behavior, or whether corruption decisions must be made before protocol execution begins. These assumptions determine what security guarantees are achievable and fundamentally shape every subsequent design decision.

\emph{Security goals} are the formal, provable properties that the system aims to achieve. These goals must be \textbf{precise}, providing unambiguous mathematical definitions that enable formal reasoning. They must be \textbf{achievable} under the stated threat model and system assumptions---goals that are theoretically impossible to achieve, or impossible under the given assumptions, are not useful goals. They must also be \textbf{meaningful} for the target application, providing real security guarantees that address the actual threats faced by users. Common security goals include safety properties such as agreement and validity, liveness properties such as termination, privacy properties that limit information leakage, and accountability properties that enable detection and attribution of misbehavior.

\emph{System assumptions} describe the environment in which the system operates, and these assumptions have profound implications for system design and applicability. Timing assumptions specify whether the system operates in a synchronous, partially synchronous, or asynchronous model. Synchronous systems assume known bounds on message delay and processing time, enabling simpler protocols but limiting applicability to environments where such bounds can be guaranteed. Authentication assumptions specify whether participants can be reliably identified and whether messages can be authenticated as coming from claimed senders. Initial setup assumptions specify whether a trusted setup phase is required, such as the generation of cryptographic keys or the establishment of shared secrets. Economic assumptions specify whether incentives are aligned, what the costs of attacks are, and whether rational adversaries would find attacks profitable.

Given the problem definition, threat model, security goals, and system assumptions, various \emph{techniques} can be employed to achieve the desired security properties. Replication strategies distribute state and computation across multiple nodes to tolerate failures. Cryptographic mechanisms, including digital signatures, threshold cryptography, zero-knowledge proofs, and secure multi-party computation, provide confidentiality, integrity, and authentication guarantees. Economic mechanisms use incentives and penalties to align participant behavior with system goals. Trusted hardware provides isolated execution environments that can be relied upon even when the rest of the system is compromised. The choice among these techniques, and the specific instantiation of chosen techniques, involves careful consideration of tradeoffs among security, performance, complexity, and trust assumptions.

\emph{Evaluation methods} demonstrate that systems work as claimed, both theoretically and practically. Formal verification uses machine-checked proofs to establish correctness with mathematical certainty. Security proofs use reduction-based arguments to show that breaking the system implies breaking underlying cryptographic or computational assumptions. Simulation testing explores system behavior under various network conditions, failure patterns, and adversary strategies. Implementation and deployment provide the ultimate validation, demonstrating that the system performs adequately under real-world conditions and that users find it valuable enough to adopt. A system that cannot be implemented or performs poorly in practice, no matter how elegant its theory, will not be adopted and therefore cannot provide security in any meaningful sense.

\subsection{Things Go Beyond Security}

While this paper focuses specifically on security properties, distributed systems must balance multiple concerns that interact with security in complex ways. \emph{Availability} ensures that the system remains operational and responsive to user requests. \emph{Resilience} enables the system to recover from failures and adapt to changing conditions. \emph{Performance} determines whether the system meets latency and throughput requirements necessary for practical use. \emph{Scalability} determines whether the system can handle growing numbers of participants, transactions, and data. \emph{Usability} affects whether users can and will employ the system correctly.

These properties often interact with security in ways that create \textbf{difficult tradeoffs}. Strong security mechanisms may reduce availability by requiring additional communication rounds or excluding potentially malicious participants. Privacy protections may hinder accountability by making it difficult to identify misbehavior. Verifiability requirements may impact performance by requiring additional cryptographic computation. The art of distributed systems design lies in finding appropriate tradeoffs for a given application domain---recognizing that there is no universally optimal solution, only solutions that are better or worse suited to particular requirements and constraints.

In this paper, we focus specifically on the security properties of agreement, consistency, privacy, verifiability and accountability. This focus allows us to examine these properties and their interactions in depth, but we acknowledge that real systems must integrate these security properties with broader operational requirements. The frameworks and fusions we discuss assume that this integration challenge can be addressed through careful engineering and appropriate architectural choices, trading off among properties when necessary while ensuring that the resulting system meets the needs of its users.

%% file: sections/key-properties.tex
\section{Key Properties}

This section examines the five fundamental properties that form the foundation of distributed security: Agreement, Consistency, Privacy, Verifiability and Accountability. We also discuss Accountability as an emerging fifth property that complements the traditional four. For each property, we provide formal definitions alongside conceptual explanations, trace the historical development of techniques for achieving that property, and discuss the current state of the art.

\subsection{Distributed Agreement: The Consensus Protocol}

\subsubsection{Foundational Concept}

Distributed agreement, commonly known as consensus, is the process by which distributed nodes reach a single version of truth despite potential malicious behavior from some participants. At its core, consensus addresses a fundamental challenge: in a distributed system where no single node can be trusted to maintain the authoritative view of system state, how can all honest nodes converge on the same view?

\begin{definition}[Byzantine Consensus]
A Byzantine consensus protocol among $n$ nodes, where up to $f$ may behave arbitrarily (Byzantine), satisfies the following properties:

\textbf{Agreement:} If any honest node decides value $v$, then all honest nodes eventually decide $v$.

\textbf{Validity:} If all honest nodes propose the same value $v$, then any honest node that decides must decide $v$.

\textbf{Termination:} All honest nodes eventually decide some value.
\end{definition}

The agreement property ensures that honest nodes never decide differently---there is no possibility of a ``split-brain'' scenario where different nodes believe different things to be true. The validity property ensures that the protocol cannot simply ignore the inputs and always decide a predetermined value; if all honest nodes want the same thing, they get it. The termination property ensures that the protocol actually produces output and cannot run forever without making progress.

These properties must hold even when up to $f$ nodes exhibit Byzantine behavior---arbitrary deviations from the protocol specification that may include sending contradictory messages to different nodes, refusing to send messages, sending malformed messages, or any other behavior an adversary might devise. The fundamental result in this space, due to Lamport, Shostak, and Pease, establishes that $n \geq 3f + 1$ is necessary and sufficient for Byzantine consensus in synchronous systems with message authentication.

\subsubsection{Classical Mechanisms}

The theoretical foundations of consensus were established through a series of seminal results that continue to shape protocol design today. Lamport's 1977 work~\cite{lamport1977proving} formalized the distinction between safety and liveness properties, providing a vocabulary that remains central to distributed systems reasoning. Safety properties, which include agreement and validity in the consensus context, are ``bad things don't happen'' guarantees that must be maintained at every point in execution. Liveness properties, which include termination, are ``good things eventually happen'' guarantees that depend critically on the fairness and timing assumptions of the system model.

The Byzantine Generals Problem~\cite{lamport1982byzantine}, published in 1982, formalized the challenge of reaching consensus with Byzantine failures. Beyond establishing the $3f+1$ bound, the paper introduced conceptual frameworks for reasoning about distributed agreement that remain influential. The generals metaphor---distinguished commanders who must coordinate their actions despite some potentially being traitors---captures the essence of Byzantine consensus in an accessible way while enabling rigorous formal analysis.

The FLP impossibility result~\cite{fischer1985impossibility}, published in 1985, demonstrated that deterministic consensus is impossible in fully asynchronous systems with even a single crash failure. The proof constructs an execution in which the protocol can remain forever undecided: messages between crucial nodes are delayed just long enough to prevent any node from being certain about the system state. Rather than ending research into practical consensus, this result channeled efforts toward partial synchrony models, where the system is asynchronous for some bounded period but eventually becomes synchronous, and toward randomized protocols that achieve consensus with probability approaching one.

Practical Byzantine Fault Tolerance (PBFT)~\cite{castro1999pbft}, introduced in 1999, marked the transition from theory to practice. PBFT achieves consensus through a three-phase protocol: pre-prepare, in which a designated leader proposes a value; prepare, in which nodes exchange messages to verify that the proposal is valid and consistent; and commit, in which nodes confirm that sufficient agreement has been reached. The protocol achieves $O(n^2)$ message complexity, which limits scalability for large networks but enables practical deployment in moderate-sized systems. Perhaps more importantly, PBFT demonstrated that Byzantine fault tolerance could achieve performance within a small constant factor of crash-fault tolerant alternatives, dispelling the notion that Byzantine resilience was too expensive for practical use.

\subsubsection{Modern Scaling}

Contemporary consensus protocols have pushed the boundaries of scalability and performance through several key innovations. Modern Byzantine fault tolerant protocols such as HotStuff~\cite{yin2019hotstuff} achieve linear $O(n)$ message complexity through threshold signatures and optimized commit structures. In HotStuff, a three-chain commit structure ensures safety even during leader changes, while threshold signatures enable compact representation of quorum certificates that would otherwise require $O(n)$ space. These advances enable Byzantine consensus to scale to hundreds or thousands of nodes, orders of magnitude beyond what PBFT could practically support.

DAG-based consensus represents another significant innovation. Rather than agreeing on a linear sequence of blocks, DAG-based systems maintain a directed acyclic graph of transactions or blocks, enabling parallel processing and higher throughput. Protocols such as Narwhal and Tusk~\cite{danezis2022narwhal} separate data dissemination from consensus ordering: a data availability layer ensures that all nodes have access to transaction data, while a consensus layer orders metadata without needing to process transaction contents. This separation enables throughput limited only by network capacity rather than by consensus overhead.

The shift toward \textbf{permissionless agreement} has arguably been the most transformative development. Bitcoin's~\cite{nakamoto2008bitcoin} proof-of-work consensus demonstrated that decentralized agreement was possible in systems where anyone could participate without registration or authorization. Rather than relying on known identities and fixed node sets, proof-of-work derives security from computational investment: participants must expend computational resources to propose blocks, and the longest valid chain represents the consensus view. Ethereum's~\cite{wood2014ethereum} proof-of-stake replaces computational work with economic stake, achieving similar security properties with dramatically lower energy consumption. These permissionless mechanisms enable consensus among parties who have no prior relationship and no reason to trust each other, yet can nevertheless cooperate to maintain consistent shared state.

\subsection{Distributed Data Consistency: The Distributed Database}

\subsubsection{Definition and Models}

Once agreement is reached on operations, distributed data consistency ensures that the resulting state is uniform and persistent across all participants. This involves both the correctness of individual operations with respect to system invariants and the integrity of the overall state across time. Different applications require different consistency guarantees, and the choice of consistency model involves fundamental tradeoffs between correctness, performance, and availability.

\begin{definition}[Consistency Models] There are three classic models. 
    
\textbf{Linearizability (Strong Consistency):} Every operation appears to take effect atomically at some point between its invocation and response. Formally, there exists a total order on all operations that is consistent with each operation's real-time ordering and with each object's sequential specification.

\textbf{Sequential Consistency:} The result of any execution is the same as if all operations were executed in some sequential order, and the operations of each individual process appear in this sequence in the order specified by its program.

\textbf{Eventual Consistency:} If no new updates are made to a given data item, eventually all accesses to that item will return the last updated value. Formally, for any read operation $r$ on object $x$ that starts after a write operation $w$ on $x$ has completed, if no other writes to $x$ occur between $w$ and $r$, then $r$ returns the value written by $w$ or a later write.
\end{definition}

Linearizability provides the strongest guarantees: operations appear to occur instantaneously at some point in real time, and all observers see the same order of operations. This enables intuitive reasoning about system behavior but requires coordination that limits availability during network partitions. Sequential consistency relaxes linearizability by requiring only a consistent global order, not necessarily the real-time order. Eventual consistency provides the weakest guarantees: it only promises that all replicas will converge to the same value eventually if updates stop, but provides no bounds on how long convergence takes or what values may be observed in the interim.

The \textbf{CAP theorem}~\cite{brewer2000cap,gilbert2002cap}, conjectured by Brewer and formally proven by Gilbert and Lynch, establishes a fundamental limitation: during a network partition, a system must choose between consistency and availability. A system that always returns responses (availability) cannot guarantee that those responses are consistent during partitions, while a system that always guarantees consistency must sometimes refuse to respond during partitions. This theorem has profound implications for system design, forcing architects to explicitly consider what tradeoffs are acceptable for their application domain.

\subsubsection{State Integrity Through Cryptography}

Cryptographic primitives play a crucial role in maintaining verifiable state in distributed databases. Hash-based data structures enable efficient verification of data integrity without requiring trust in the party storing the data.

\begin{definition}[Merkle Tree]
A Merkle tree is a hash-based data structure where each leaf node contains the hash of a data block, and each non-leaf node contains the hash of its children. For a tree with root hash $h_{root}$, a membership proof for data block $d$ consists of the sibling hashes along the path from $d$'s leaf to the root. The verifier computes:
\[
h_{root} \stackrel{?}{=} H(\ldots H(H(d), sibling_1) \ldots, sibling_k)
\]
where $H$ is the hash function and $sibling_i$ are the provided sibling hashes.
\end{definition}

Merkle trees provide tamper-evidence: any modification to the underlying data changes the leaf hash, which propagates up the tree to change the root hash. Membership proofs are logarithmic in the number of data blocks, enabling efficient verification even for large datasets. This property is essential for distributed databases where participants must verify that their view of the data is consistent with others without downloading the entire database.

Authenticated data structures extend this concept beyond simple trees. Verkle trees, for example, provide more compact proofs through vector commitments, reducing proof size from logarithmic to polylogarithmic in the number of elements. Accumulators provide constant-size membership proofs for sets, enabling efficient verification of membership or non-membership. These structures are increasingly important as distributed databases grow to petabyte scale, where traditional Merkle proofs would be too large for practical use.

\subsection{Distributed Privacy: Secure Multi-Party Computation}

\subsubsection{Theoretical Foundations}

Secure Multi-Party Computation (MPC) addresses a fundamental challenge: how can multiple parties compute a function over their private inputs without revealing those inputs to each other? The problem was formalized by Yao in 1982~\cite{yao1982protocols} through the Millionaires' Problem: two millionaires wish to determine who is richer without revealing their actual wealth to each other. This seemingly simple problem encapsulates the core challenge of privacy-preserving distributed computation.

\begin{definition}[Secure Multi-Party Computation]
Let $f: (\{0,1\}^*)^n \rightarrow (\{0,1\}^*)^n$ be a function, and let $P_1, \ldots, P_n$ be parties with private inputs $x_1, \ldots, x_n$. An $n$-party protocol $\pi$ securely computes $f$ if for any probabilistic polynomial-time adversary $\mathcal{A}$ corrupting a subset of parties $I \subset \{1, \ldots, n\}$, there exists a simulator $\mathcal{S}$ such that:

\[
\{\textsf{Real}_{\pi, \mathcal{A}(z)}(\vec{x})\}_{\vec{x}, z} \approx \{\textsf{Ideal}_{f, \mathcal{S}(z)}(\vec{x})\}_{\vec{x}, z}
\]

where $\textsf{Real}$ denotes the view of the adversary in the real protocol execution, and $\textsf{Ideal}$ denotes the view in an ideal world where a trusted party computes $f$ and provides outputs.
\end{definition}

The simulation paradigm underlying this definition captures the intuition that the adversary should learn nothing from the protocol beyond what is implied by its own input and output. If the adversary's view in the real protocol can be simulated given only what the adversary would learn in an ideal world with a trusted party, then the protocol leaks no additional information. This definition provides composability: protocols proven secure under this definition remain secure when composed with other secure protocols.

\subsubsection{Techniques for Privacy-Preserving Execution}

Several fundamental techniques enable privacy-preserving distributed computation, each with distinct tradeoffs.

Garbled circuits, introduced by Yao, allow two parties to compute any Boolean circuit while keeping their inputs private. The protocol works by having one party (the garbler) encrypt the truth table of each gate in the circuit, while the other party (the evaluator) obtains the appropriate encryption keys through oblivious transfer without learning which keys correspond to which input values. The evaluator then computes the encrypted circuit gate by gate, eventually obtaining the encrypted output, which both parties can then decrypt. Modern optimizations, including free XOR gates, half gates, and batched oblivious transfer, have reduced the overhead of garbled circuits by orders of magnitude, enabling practical computation of complex functions.

Secret sharing provides an alternative approach based on distributing information among multiple parties such that no individual party can learn anything about the secret. Shamir's secret sharing~\cite{shamir1979share} splits a secret $s$ into $n$ shares by choosing a random polynomial $p(x)$ of degree $t-1$ with $p(0) = s$ and setting share $i$ to $p(i)$. Any $t$ shares can reconstruct the secret through polynomial interpolation, while fewer than $t$ shares reveal no information. This threshold structure enables secure computation: parties can perform arithmetic operations on secret-shared values locally, and the result remains secret-shared. Multiplication of secret-shared values requires communication, but efficient protocols minimize this overhead.

Homomorphic encryption allows computation on encrypted data without decryption. Additively homomorphic schemes, such as Paillier encryption, support addition of encrypted values, enabling applications like encrypted voting and private aggregation. \textbf{Fully homomorphic encryption} (FHE), realized by Gentry in 2009~\cite{gentry2009fhe}, supports arbitrary computation on encrypted data, enabling a party to compute any function on encrypted inputs and produce an encrypted result that only the secret key holder can decrypt. While FHE remains computationally expensive, ongoing improvements in efficiency have enabled practical applications in privacy-preserving machine learning and secure database queries.

\subsubsection{State of the Art and Related Approaches}

Modern MPC implementations often combine multiple techniques to balance security and performance. Hybrid protocols might use garbled circuits for Boolean operations and secret sharing for arithmetic, leveraging the strengths of each approach. Preprocessing phases can shift expensive cryptographic operations offline, enabling efficient online computation. The state of the art achieves throughput of millions of operations per second for simple computations, making MPC practical for an expanding range of applications.

Trusted Execution Environments (TEEs)~\cite{costan2016sgx} provide an alternative approach based on hardware isolation. Intel SGX, ARM TrustZone, and similar technologies provide secure enclaves where sensitive computation can occur, protected from the operating system and other software on the host machine. While TEEs offer better performance than pure cryptographic approaches, they require trusting the hardware manufacturer and have been subject to side-channel attacks. TEEs are best viewed as complementary to cryptographic MPC, providing practical solutions for applications where the threat model is compatible with hardware trust assumptions.

Local differential privacy~\cite{dwork2006dp} and federated learning~\cite{mcmahan2017fl} address related privacy concerns through statistical rather than cryptographic approaches. Local differential privacy adds noise at the data source, providing mathematical bounds on what can be inferred about individual inputs. Federated learning enables model training without centralizing data, with participants computing gradients locally and only sharing aggregated updates. These techniques provide scalable privacy protection for machine learning and analytics applications, though they offer weaker guarantees than MPC: privacy is statistical rather than perfect, and the computed function is approximate rather than exact.

\subsection{Distributed Verifiability: Zero-Knowledge Proofs and Accumulators}

\subsubsection{The Concept of Verifiability}

Distributed verifiability enables participants to prove the correctness of computations without requiring others to re-execute them. This capability is essential in distributed systems for several reasons. First, computation may be expensive, and requiring all participants to re-execute would be prohibitively costly. Second, the prover may have access to data that verifiers do not, making re-execution impossible. Third, trust must often be established between parties who have no prior relationship and no reason to trust each other's assertions. Zero-knowledge proofs provide a powerful mechanism for addressing all three challenges.

\begin{definition}[Zero-Knowledge Proof]
A zero-knowledge proof system for language $L$ consists of algorithms $(P, V)$ where $P$ is the prover and $V$ is the verifier. For any instance $x$ and witness $w$:

\textbf{Completeness:} If $(x, w) \in R_L$ (where $R_L$ is the relation defining $L$), then $\Pr[V(P(x, w)) = 1] = 1$.

\textbf{Soundness:} If $x \notin L$, then for any malicious prover $P^*$, $\Pr[V(P^*(x)) = 1] \leq \text{negl}(\lambda)$.

\textbf{Zero-Knowledge:} There exists a simulator $S$ such that for all $(x, w) \in R_L$, the view of $V$ in an honest execution is computationally indistinguishable from $S(x)$.
\end{definition}

Completeness ensures that honest provers can always convince verifiers of true statements. Soundness ensures that malicious provers cannot convince verifiers of false statements except with negligible probability. Zero-knowledge ensures that the proof reveals nothing beyond the statement's validity---the verifier learns that the statement is true but nothing else about the witness or private inputs.

\subsubsection{The Efficiency Revolution}

Modern zero-knowledge proof systems have undergone an efficiency revolution that has transformed them from theoretical constructs to practical tools. The key developments include:

Succinct Non-interactive ARguments of Knowledge (zk-SNARKs) provide constant-size proofs and linear-time verification. A zk-SNARK proof for arbitrary computation produces a proof of a few hundred bytes that can be verified in milliseconds, regardless of the computation's complexity. This succinctness enables applications where verifiers are resource-constrained or where proofs must be transmitted over limited bandwidth. The tradeoff is a trusted setup: a ceremony that generates public parameters also creates a toxic waste that, if not destroyed, could be used to create false proofs. Techniques for multiparty computation-based setup ceremonies and transparent alternatives have mitigated this concern for many applications.

Scalable Transparent ARguments of Knowledge (zk-STARKs) eliminate the trusted setup requirement while maintaining succinct verification. zk-STARKs rely only on collision-resistant hash functions, providing security against quantum adversaries. The tradeoff is larger proof sizes---tens to hundreds of kilobytes rather than hundreds of bytes---though ongoing improvements continue to narrow this gap.

\textbf{Bulletproofs}~\cite{bunz2018bulletproofs} provide efficient range proofs without trusted setup, enabling confidential transactions in blockchain systems. A Bulletproof proving that a value lies in a 64-bit range produces a proof of only a few kilobytes, with verification time logarithmic in the range size. The combination of reasonable proof sizes, no trusted setup, and standard cryptographic assumptions has made Bulletproofs popular for privacy-preserving cryptocurrency applications.

\subsubsection{Cryptographic Accumulators}

Cryptographic accumulators provide compact representations of sets supporting efficient membership proofs. Like Merkle trees, accumulators enable verification of set membership without revealing the entire set; unlike Merkle trees, accumulators can provide constant-size membership proofs.

\begin{definition}[Cryptographic Accumulator]
A cryptographic accumulator for a set $S$ consists of algorithms (Setup, Accumulate, Prove, Verify):

\textbf{Setup($1^\lambda$) $\rightarrow$ (pk, vk):} Generates public and verification keys.

\textbf{Accumulate(pk, $S$) $\rightarrow$ acc:} Computes the accumulator value for set $S$.

\textbf{Prove(pk, $S$, $x$) $\rightarrow$ $\pi$:} Generates a membership proof for $x \in S$.

\textbf{Verify(vk, acc, $x$, $\pi$) $\rightarrow$ $\{0,1\}$:} Verifies that $x$ is in the set accumulated as acc.
\end{definition}

RSA accumulators use the RSA assumption to provide constant-size membership proofs and efficient batch updates. The accumulator value is $g^{\prod_{p \in S} p} \mod N$ where $N$ is an RSA modulus and $p$ ranges over primes representing set elements. Membership proofs are single group elements, and verification requires a single modular exponentiation. The tradeoff is a trusted setup to generate the RSA modulus, though techniques for multiparty computation-based setup can mitigate this requirement.

Bilinear accumulators use pairings on elliptic curves to enable both membership and non-membership proofs with efficient verification. Vector commitments generalize accumulators to support position-based proofs, enabling efficient updates to committed data structures. These primitives are essential building blocks for verifiable databases, stateless blockchains, and other systems requiring efficient proofs about large datasets.

\subsection{Accountability}

\subsubsection{The Role of Accountability}

Accountability is an emerging property that complements traditional security properties in an important way. While agreement, consistency, privacy, and verifiability are primarily concerned with preventing bad behavior, accountability is concerned with detecting, attributing, and punishing bad behavior that occurs despite preventive measures. In many practical settings, accountability provides a crucial layer of defense: even if a system cannot technically prevent all attacks, the threat of detection and punishment may deter potential attackers.

\begin{definition}[Accountable System]
An accountable distributed system satisfies:

\textbf{Detection:} If any participant misbehaves by deviating from the protocol specification, there exists a procedure that can identify that misbehavior has occurred.

\textbf{Attribution:} Given detected misbehavior, the responsible participant(s) can be identified with cryptographic certainty.

\textbf{Evidence:} The detection and attribution process produces cryptographic evidence that can be verified by any third party.

\textbf{Punishment:} There exist mechanisms to impose consequences on identified misbehaving parties, such as economic penalties or reputation damage.
\end{definition}

The detection property ensures that misbehavior does not go unnoticed. The attribution property ensures that when misbehavior is detected, the guilty parties can be identified rather than misbehavior being attributed to the wrong participants or to generic system failures. The evidence property ensures that accountability claims are verifiable rather than requiring trust in the accuser. The punishment property ensures that accountability has teeth---that there are real consequences for detected misbehavior.

\subsubsection{Techniques for Accountability}

Digital signatures provide the foundation for accountability through non-repudiation. When a participant signs a message, the signature provides cryptographic evidence that the participant authored or approved that message. In consensus protocols, signed messages enable attribution of protocol violations: if a node signs two conflicting blocks at the same height, the signatures constitute undeniable evidence of misbehavior.

Audit logs extend accountability by maintaining tamper-evident records of system actions. Authenticated data structures such as Merkle trees enable the construction of append-only logs where entries cannot be modified or deleted without detection. Certificate transparency logs, for example, use Merkle trees to create publicly auditable records of all certificates issued by certificate authorities, enabling detection of misissuance.

Economic mechanisms provide accountability through incentives rather than purely technical means. In proof-of-stake systems, validators who misbehave can have their stake slashed---confiscated as punishment for protocol violations. This creates direct financial consequences for misbehavior, transforming accountability from a technical property into an economic deterrent. The threat of slashing aligns validator incentives with honest behavior even when technical prevention is imperfect.

The combination of verifiability and accountability is particularly powerful. Accountable safety, as implemented in protocols like Casper FFG~\cite{buterin2017casper}, ensures that if a safety violation occurs, a substantial fraction of validators can be proven to have misbehaved and have their stake slashed. This transforms the security model: attacks remain possible in principle, but become economically irrational because the expected cost (probability of detection times stake slashed) exceeds any potential gain. Accountability thus provides security through deterrence where technical prevention is insufficient or too expensive.

%% file: sections/frontier-fusions.tex
\section{The Frontier: Fusions}

The properties discussed in the previous section have historically been studied and implemented in relative isolation by different research communities. The \textbf{frontier} of distributed security research lies at the intersection of these properties, where their combination creates new capabilities that neither property could provide alone. This section examines how these fusions are reshaping the landscape of distributed security, creating new paradigms that transcend traditional category boundaries.

\subsection{Fusion of Agreement and Data Consistency: The Blockchain Technique}

The blockchain represents perhaps the \emph{most significant fusion} in distributed security history: the combination of consensus protocols with distributed database systems. Traditional distributed systems maintained a clear separation between these concerns. Consensus protocols like Paxos~\cite{lamport1998paxos} and Raft~\cite{ongaro2014raft} were designed to determine the order of operations across a set of nodes, while database systems managed the state resulting from those operations. The blockchain paradigm fuses these into a single, inseparable construct.

The key insight underlying this fusion is that the data structure for recording state can itself serve as the mechanism for achieving and encoding agreement. In a blockchain, each block contains both transaction data (the database aspect) and a cryptographic reference to the previous block (creating the ordering). This chained structure means that the process of agreeing on the next block simultaneously determines both the order of new transactions and the resulting state of the system. The two concerns that were previously separate become inextricably linked.

This fusion has profound implications for security. Because each block cryptographically commits to all previous blocks, the history becomes tamper-evident: modifying any historical block would change its hash, invalidating all subsequent blocks. Because consensus determines block order, all honest participants agree on the canonical history. The combination creates what has been called a ``truth machine''---a system where anyone can verify the complete history of transactions, the state can be reconstructed at any point in time, and no trusted authority is required to maintain correctness.

Bitcoin~\cite{nakamoto2008bitcoin} demonstrated this fusion's power in a permissionless setting, where anyone could participate without registration or authorization. The proof-of-work consensus mechanism provided agreement among anonymous participants with no prior trust relationships, while the blockchain data structure provided consistent state. Ethereum~\cite{wood2014ethereum} extended this vision by adding programmable smart contracts---general-purpose code that executes on the blockchain---enabling complex decentralized applications beyond simple value transfer. The success of these systems has inspired numerous variations optimizing for different properties: faster consensus, better scalability, enhanced privacy, or improved energy efficiency.

The blockchain paradigm has also revealed challenges unique to this fusion. The tight coupling of agreement and state means that the blockchain must record all transactions, creating scalability challenges as history grows indefinitely. The need for all nodes to verify all transactions limits throughput compared to traditional databases. Various layer-2 solutions and sharding approaches address these challenges by moving some computation off-chain while maintaining the security guarantees of the underlying blockchain, but these solutions introduce their own complexity and tradeoffs.

\subsection{Fusion of Consistency and Verifiability: Succinct Distributed Proof}

The combination of verifiability primitives, particularly zero-knowledge proofs, with distributed databases enables a paradigm shift in how participants can verify system state. Rather than downloading and verifying entire databases, participants can verify compact cryptographic proofs that attest to state correctness. This fusion addresses one of the fundamental scalability challenges in distributed systems: how can participants with limited resources verify the correctness of large-scale state?

The key insight is that zero-knowledge proofs can compress the entire history of state transitions into a single, compact proof. Rather than verifying each transaction individually, a verifier can check a proof that attests to the correct execution of all transactions from some known starting state to the current state. This proof can be constant or logarithmic in size regardless of the number of transactions, enabling efficient verification even for systems with billions of historical transactions.

\textbf{Zero-knowledge rollups} exemplify this fusion in practice. In a zk-rollup, transactions are processed off-chain in batches, and a zero-knowledge proof is generated attesting to the correct execution of all transactions in the batch. This proof is then submitted to and verified on the main chain. The on-chain verification is constant-time regardless of the number of transactions in the batch, enabling throughput limited only by the proof generation capacity rather than by on-chain verification constraints. Systems like zkSync and StarkNet demonstrate that this approach can achieve thousands of transactions per second while maintaining the security guarantees of the underlying blockchain.

The fusion also enables new participant models. Stateless clients can participate in networks without maintaining full state, verifying only the proofs relevant to their interests. Light clients can verify the correctness of specific account balances or transaction inclusions without processing the entire chain. This dramatically lowers the resource requirements for participation, potentially enabling blockchain verification on mobile devices or embedded systems that could never run full nodes.

Beyond blockchains, the fusion of consistency and verifiability enables new models for distributed databases. Verifiable databases allow clients to query data and verify that responses are consistent with the committed state, without trusting the database operator. Techniques like query authentication and authenticated indexing enable efficient proofs of query results, while accumulator-based commitments enable compact representations of large datasets. These capabilities are essential for applications requiring trustless data access, from decentralized identity systems to auditable machine learning.

\subsection{Fusion of Privacy and Verifiability: Collaborative Zero-Knowledge}

The combination of multi-party computation with zero-knowledge proofs represents one of the \textbf{most sophisticated fusions} in distributed security. This integration enables what has been called \emph{collaborative zero-knowledge} (C-ZK): multiple parties can collaboratively generate a zero-knowledge proof about their collective private data without any single party revealing their individual inputs or learning the complete witness~\cite{ozdemir2022collaborative}.

The challenge this fusion addresses is fundamental. Traditional zero-knowledge proofs assume a single prover who knows the complete witness---all the private data that the proof is about. But in many distributed applications, no single party has or should have access to all the relevant data. Consider a financial audit: multiple institutions might need to prove that their combined operations satisfy certain regulatory requirements, without any institution revealing its individual data to the others or to the auditor.

Collaborative zero-knowledge solves this through a combination of secret sharing and proof generation. The witness is distributed across multiple parties using secret sharing, ensuring that no individual party can reconstruct it. The parties then engage in a distributed protocol to generate the zero-knowledge proof, performing the cryptographic operations on their secret-shared portions of the witness. The resulting proof is publicly verifiable like any zero-knowledge proof, but no party learns anything beyond what the proof statement itself reveals.

This fusion has profound implications for privacy-preserving verification. It enables systems that are simultaneously publicly verifiable and privately computed, with no trusted third party required to aggregate private data. Applications include private smart contracts that execute on hidden data while proving correct execution, collaborative fraud detection that identifies suspicious patterns across institutions without revealing individual transactions, and privacy-preserving audits that verify regulatory compliance without exposing underlying business data.

The technical challenges of this fusion are significant. Distributed proof generation requires careful protocol design to ensure that the proof is generated correctly even if some parties misbehave. The computational overhead of combining MPC with ZKP is substantial, though ongoing improvements in both areas continue to narrow the gap with non-private alternatives. The communication patterns required for distributed proof generation may not align well with network topologies, requiring careful optimization for practical deployment.

\subsection{Fusion of Verifiability and Accountability: Economic Security}

The combination of verifiability mechanisms with accountability creates systems where misbehavior is not only detectable but attributable with concrete consequences. This fusion is particularly powerful in economic systems where security can be achieved through the credible threat of punishment rather than purely through technical prevention.

The key insight is that cryptographic proofs can serve not just to verify correct behavior but to provide evidence of misbehavior. When all protocol messages are signed and all state transitions are recorded in authenticated data structures, any deviation from the protocol specification produces cryptographic evidence that can be used for accountability. This evidence is non-repudiable---the misbehaving party cannot deny their actions without denying the validity of their signatures.

\textbf{Proof-of-stake systems} exemplify this fusion. In protocols like Ethereum's proof-of-stake~\cite{buterin2020ethereum}, validators are required to sign blocks and attestations, creating a cryptographic record of their actions. If a validator signs two conflicting blocks at the same height, the signatures constitute undeniable evidence of misbehavior. The protocol can then slash the validator's stake---confiscating it as punishment for the protocol violation. This creates direct financial consequences for misbehavior, aligning validator incentives with honest behavior.

The fusion transforms the security model in an important way. Traditional security analysis focuses on what adversaries can and cannot do: can they violate safety properties? can they prevent liveness? The accountability fusion adds a new dimension: even if adversaries can technically violate security properties, doing so may be economically irrational. The expected cost of misbehavior---the probability of detection times the penalty if detected---may exceed any potential gain. This deterrence-based security complements technical prevention, particularly in settings where perfect prevention is impossible or too expensive.

Beyond slashing for consensus violations, this fusion enables broader accountability mechanisms. Fraud proofs allow anyone to demonstrate that a state transition was incorrect, earning a reward while penalizing the fraudulent operator. Optimistic rollups rely on this mechanism for security: state updates are assumed correct unless challenged, but anyone can challenge and prove fraud, creating accountability without requiring all participants to verify all transactions. Dispute resolution mechanisms in prediction markets and decentralized arbitration systems use similar principles to provide accountability in complex multi-party interactions.

\subsection{Emerging Fusions and Triple Combinations}

The fusions discussed above represent pairwise combinations of properties, but the most ambitious systems combine three or more properties. These triple combinations create capabilities that would be impossible with fewer properties, though they also present the greatest technical challenges.

Private blockchains combine agreement, consistency, and privacy to enable decentralized systems where participants can transact without revealing their activities to all network participants. Techniques like confidential transactions hide transaction amounts, while zero-knowledge proofs enable verification that hidden transactions are valid. More sophisticated systems use secure multi-party computation to enable smart contracts that operate on encrypted data, achieving privacy for both transaction contents and contract execution. These systems enable enterprise applications that require both the trust guarantees of decentralized consensus and the confidentiality of traditional business systems.

Verifiable multi-party computation combines privacy with verifiability, enabling public verification of private computations. A group of parties can compute a function over their private inputs and produce not just the output but a proof that the computation was performed correctly. This enables applications like privacy-preserving audits, where an auditor can verify that a computation satisfies certain properties without learning the private inputs, or collaborative analytics, where multiple organizations can prove statistical properties of their combined data without revealing individual records.

Accountable privacy systems represent a particularly interesting combination, preserving privacy under normal operation while enabling accountability when warranted. These systems might use threshold cryptography where private keys are distributed among multiple parties, enabling decryption only when a sufficient number agree that conditions for accountability have been met. They might use verifiable encryption to prove that encrypted data satisfies certain properties, enabling courts or other authorized parties to decrypt under specific circumstances. The challenge is designing systems that provide meaningful privacy guarantees while enabling appropriate accountability---a balance that has proven difficult to achieve in practice.

Several other fusions are emerging at the research frontier. Private consensus protocols aim to achieve agreement on encrypted or hidden values, enabling applications where even the consensus participants should not learn transaction contents. Encrypted databases maintain consistency while keeping data encrypted from the database operators themselves. Cross-chain protocols enable fusions across different blockchain systems, combining properties from each. These emerging fusions suggest that the combinatorial space of property combinations remains largely unexplored, with significant research opportunities in systematically exploring and optimizing these combinations.

%% file: sections/future-challenges.tex
\section{Future Challenges and Research Vision}

Having surveyed the current state of distributed security and its emerging fusions, we now turn to the challenges and opportunities that will shape the field's future. This section outlines key research directions organized around three fundamental questions: 
\begin{itemize}
    \item How do we discover new security properties that may be important for emerging applications?
    \item How do we systematically combine multiple properties in ways that create synergies rather than conflicts?
    \item How do we understand and manage the complex interactions between properties when they are combined?
\end{itemize}

\subsection{Discovering New Properties}

The properties we have discussed emerged from decades of research and practice. Yet there may be other fundamental properties that remain undiscovered, properties that will become essential as distributed systems evolve into new domains and face new challenges. The question of how to systematically discover new security properties is itself an important research challenge.

Historically, properties have been discovered through two primary mechanisms, each with its own strengths and limitations. Application-driven discovery occurs when new applications expose requirements that existing security frameworks cannot adequately express. The rise of blockchain systems, for instance, revealed that traditional consistency models were insufficient to capture the requirements of systems where consistency must be verifiable by arbitrary third parties. The emergence of decentralized finance revealed that accountability---the ability to identify and punish misbehavior---was essential in economic systems, even though accountability had received relatively little attention in traditional distributed systems research.

Theoretical discovery, by contrast, emerges from fundamental research often without immediate application in mind. Zero-knowledge proofs were a theoretical construct for nearly two decades before finding practical application in blockchain systems. The study of composability in cryptographic protocols revealed fundamental limitations that have shaped how we think about protocol design. These theoretical discoveries often provide the vocabulary and frameworks that later prove essential for addressing practical challenges, even when the original motivation was purely intellectual curiosity.

A systematic approach to discovering new properties should draw on both mechanisms. Researchers should closely examine emerging applications---AI systems that require verifiable training data, IoT networks that must maintain security despite device compromises, quantum communication systems that rely on fundamentally different physical principles---to identify security requirements not captured by existing frameworks. Cross-domain inspiration from fields like game theory, mechanism design, information theory, and formal methods may reveal properties applicable to distributed systems that have not been considered in the distributed systems literature. Formal exploration using rigorous mathematical methods can systematically explore the space of possible security properties, identifying gaps in current coverage.

Several candidates for new properties warrant particular attention. Composability addresses how security guarantees compose when systems interact---a crucial concern as distributed systems become increasingly interconnected. Upgradability addresses how systems maintain security properties while evolving over time, particularly when upgrades may require coordination among mutually distrusting parties. Censorship resistance addresses the guarantee that no participant or coalition can prevent others from participating in or benefiting from the system---a property distinct from both availability and agreement. Each of these represents a security concern that is not fully captured by existing properties, and each may become increasingly important as distributed systems evolve.

\subsection{Converging Multiple Properties}

With five fundamental properties, there are 26 possible combinations of two or more properties ($2^5 - 5 - 1 = 26$). Many of these combinations remain largely unexplored, and even the combinations that have been studied lack systematic frameworks for reasoning about tradeoffs and optimizations. The challenge of converging multiple properties is to develop principled approaches for combining properties in ways that maximize benefit while managing cost.

Several principles appear to govern successful property convergence. Shared assumptions facilitate combination: properties that rely on compatible trust models, network assumptions, and cryptographic primitives can be combined more easily than properties that make conflicting assumptions. Complementary overhead enables efficiency: when the computational costs of different properties overlap---when, for instance, the cryptographic operations for privacy can be amortized with those for verifiability---the combined system can be more efficient than two separate systems. Semantic compatibility is essential: properties must be expressible within a common formal framework to enable rigorous reasoning about their combination.

The development of \textbf{universal composability frameworks} represents an important research direction. Such frameworks would specify conditions under which protocols satisfying individual security properties can be composed to produce systems satisfying combined properties, without requiring case-by-case security proofs for each combination. The universal composability framework~\cite{canetti2001uc} developed in the cryptography community provides a model, but extending such frameworks to encompass all the properties relevant to distributed systems---including agreement and consistency, which have not traditionally been studied within cryptographic frameworks---remains an open challenge.

Property-aware design methodologies would consider multiple security properties from initial system design rather than attempting to add properties as afterthoughts. Such methodologies would provide guidance on how architectural decisions affect multiple properties simultaneously, enabling designers to identify opportunities for synergy and avoid architectural choices that preclude important combinations. The development of design patterns for secure distributed systems---reusable architectural templates that provide known security properties---would make integrated security more accessible to practitioners.

Systematic tradeoff analysis would quantify the relationships between properties, enabling optimization under constraints. If we understand the ``exchange rate'' between properties---how much performance must be sacrificed for additional privacy, how much latency must be tolerated for additional verifiability---we can make principled decisions about what combinations to pursue for particular applications. This requires not just theoretical analysis but empirical measurement of real systems to calibrate models and identify optimization opportunities.

\subsection{Understanding Synergies}

Even when properties can be combined, understanding their interactions remains challenging. Properties may reinforce each other, as when accountability strengthens agreement by deterring Byzantine behavior. They may create tension, as when strong privacy conflicts with accountability by making it difficult to identify misbehavior. They may produce emergent behaviors, as when the combination of agreement and verifiability in blockchain systems created censorship resistance---a property that was not explicitly designed but emerged from the combination.

Managing computational overhead is perhaps the most pressing practical challenge. The ``cryptographic tax'' of combining MPC and ZKP with consensus protocols can be substantial. General multi-party computation remains orders of magnitude slower than plaintext computation, raising the question of whether MPC can be made practical for high-throughput consensus systems. Zero-knowledge proof generation introduces latency that may be incompatible with the low-latency requirements of some consensus applications. Hardware acceleration through specialized circuits for cryptographic operations may help, but systematic approaches to overhead management---architectural patterns, protocol optimizations, and resource allocation strategies---remain important research challenges.

A theory of synergies would formalize our understanding of property interactions. Such a theory would quantify the relationships between properties, enabling prediction of emergent behaviors and optimization of combinations under constraints. Given requirements for throughput, latency, and cost, what is the optimal combination of properties? What are the fundamental limits on what combinations can achieve? Can we predict when combinations will produce beneficial emergence versus harmful conflicts? These questions represent fundamental research challenges that bridge cryptography, distributed systems, and formal methods.

The tension between privacy and accountability exemplifies the challenges of understanding synergies. In isolation, both properties are well-understood, and techniques for achieving each individually are mature. But combining them is fundamentally difficult: privacy requires hiding information, while accountability requires revealing information when misbehavior occurs. Resolving this tension requires careful design, potentially using threshold mechanisms that preserve privacy under normal operation but enable disclosure when sufficient evidence of misbehavior exists, or using zero-knowledge proofs to enable verification without revealing underlying data. The development of general frameworks for managing such tensions would advance the field significantly.

\subsection{Post-Quantum Distributed Security}

The emergence of quantum computing threatens many cryptographic primitives underlying distributed security. Shor's algorithm can break RSA and elliptic curve cryptography, which underpin most digital signatures and key exchange protocols. Grover's algorithm reduces the security of symmetric cryptography and hash functions by approximately half, requiring larger key sizes to maintain security levels. The transition to post-quantum cryptography will require rethinking many aspects of distributed security.

The challenge extends beyond simply replacing cryptographic primitives. Post-quantum signatures are larger and slower than their classical counterparts, affecting the efficiency of consensus protocols that rely heavily on signatures. Post-quantum key exchange has different security properties and performance characteristics than Diffie-Hellman, affecting the design of secure channels. Some post-quantum assumptions are less well-studied than classical assumptions, requiring careful analysis of the resulting security guarantees.

Research directions include developing consensus protocols optimized for post-quantum signatures, designing zero-knowledge proof systems based on post-quantum assumptions, and exploring how quantum computing might enhance rather than threaten distributed security through techniques like quantum key distribution and quantum random number generation. The transition to post-quantum distributed security will likely span decades, requiring careful attention to backwards compatibility and migration strategies.

\subsection{Human Factors in Distributed Security}

Security systems ultimately serve human users, yet human factors are often neglected in distributed security research. This neglect has practical consequences: systems that are theoretically secure may fail in practice because users misunderstand their security properties, make errors in configuration or operation, or simply choose not to use them.

Users often have incorrect mental models of distributed systems, leading to security failures. The distinction between ``trustless'' and ``minimizing trust'' may be lost on non-technical users, leading them to place more faith in system outputs than is warranted. The implications of key management---that losing a private key means losing access forever---may not be fully appreciated until it is too late. The relationship between privacy properties and practical privacy---the difference between cryptographic privacy and operational privacy---may be misunderstood, leading users to reveal information through side channels.

Strong security often conflicts with usability. Key management remains one of the most significant barriers to adoption of secure systems. Recovery mechanisms that provide convenience may undermine security guarantees. The complexity of understanding what a system does and does not guarantee may overwhelm users, leading them to ignore security properties entirely.

Research directions include designing usable security mechanisms that enable users to perform security-critical operations correctly without requiring deep understanding of underlying cryptography, making security properties visible and comprehensible through appropriate abstractions and visualizations, and developing trust interfaces that help users make informed decisions about what and whom to trust based on system properties rather than marketing claims. These challenges require collaboration between security researchers and human-computer interaction specialists, bridging communities that have historically had limited interaction.

%% file: sections/conclusion.tex
\section{Conclusion}

Over the past forty-five years, distributed security has evolved from foundational theory to global-scale decentralized systems. Early breakthroughs such as the safety–liveness framework, the Byzantine Generals Problem, and impossibility results established the fundamental limits of consensus and shaped the intellectual core of the field. Subsequent advances transformed theory into practice: crash-fault tolerant protocols like Paxos and Raft, Practical Byzantine Fault Tolerance, and globally distributed databases demonstrated that reliable and secure agreement was achievable in real systems. The advent of blockchain systems, beginning with Bitcoin and later Ethereum, marked a new phase—showing that cryptography, incentives, and distributed protocols could enable trust without centralized authorities. Across this progression, theory, systems engineering, cryptography, and economics continuously reinforced one another, expanding both the capabilities and the ambitions of distributed systems.

Today, the frontier lies not within isolated research silos but at their intersections. Modern applications demand agreement, consistency, privacy, verifiability, and accountability simultaneously, pushing us toward what we call Architectural Synthesis: designing systems where multiple security properties emerge from a unified foundation rather than being layered independently. In such architectures, consensus, storage, and cryptography are deeply intertwined, and trust arises from structural design rather than external enforcement. Realizing this vision requires cross-disciplinary collaboration, principled reasoning about property composition, and close integration of theory with practical deployment. The future of distributed security depends on breaking traditional boundaries and systematically engineering synergistic systems that enable cooperation without requiring blind trust.